\documentclass[journal]{IEEEtran}
\usepackage[cmex10]{amsmath}
\usepackage{enumerate}

\usepackage{amsfonts,amscd}
\usepackage{epsfig}
\usepackage{float}

\newtheorem{theorem}{Theorem}[section]
          
\newtheorem{lemma}[theorem]{Lemma}              
\newtheorem{proposition}[theorem]{Proposition}

\newcommand{\be}{\begin{equation}}
\newcommand{\ee}{\end{equation}}
\newcommand{\bea}{\begin{eqnarray}}
\newcommand{\eea}{\end{eqnarray}}
\newcommand{\beann}{\begin{eqnarray*}}
\newcommand{\eeann}{\end{eqnarray*}}
\newcommand{\ip}[2]{\langle{#1}|{#2}\rangle}

\begin{document}
\title{Entropy rate calculations of algebraic measures}

\author {Katy~Marchand,
        Jaideep~Mulherkar,
        and~Bruno~Nachtergaele
\thanks{Katy Marchand is currently unaffiliated.}%
\thanks{Jaideep Mulherkar is with DA institute of information and communication technology, Gandhinagar, India, email:jaideep\_mulherkar@daiict.ac.in.}%
\thanks{Bruno Nachtergaele is with the Department of Mathematics University of California, Davis USA e-mail:bxn@math.ucdavis.edu.}}

\maketitle

\begin{abstract}
Let $K = \{0,1,...,q-1\}$. We use a special class of translation invariant
measures on $K^\mathbb{Z}$ called  algebraic measures to study the 
entropy rate of a hidden Markov processes. Under some irreducibility assumptions
of the Markov transition matrix we derive exact formulas for the entropy rate of
a general  $q$  state hidden Markov process derived from a Markov source
corrupted by a specific noise model. We obtain upper bounds on the error when using an approximation 
to the formulas and numerically compute the entropy rates of two and three state hidden Markov models.
\end{abstract}


\section{Introduction}
\IEEEPARstart{I}n this paper we study the entropy rate of a hidden Markov process using a class of translation-invariant measures 
on the chain $K^{\mathbb{Z}}$ where $K=\{0,....,q-1\}$. These measures known as manifestly positive algebraic measures and their properties were
first studied in \cite{FNS1992}. A one to one correspondence was shown in \cite{FNS1992} a between manifestly positive algebraic measures and
hidden Markov process. We use the properties of the algebraic measures to give formulas for the entropy rate of a hidden Markov
process derived from a certain noise model that we will describe later.

In information theory one models an information source as a stochastic process 
$\{X_i\}_{i=1}^\infty$ with each $X_i$ a random
variable taking values in the alphabet set $K=\{0,...,q-1\}$. 
The Shannon entropy of a random variable $X$ taking values in a set $K$ is defined as
\begin{equation}
\label{eq:Entropy}
S(X) = -\sum_{x\in K} p(x) \log_2 p(x)
\end{equation}
The entropy of the the source for the first $n$ transmitted symbols is given by the joint entropy of $X_1,...,X_n$
\begin{equation}
\label{eq:JointEntropy}
S_n(X_1,X_2,...,X_n) = -\sum_{x_1,...,x_n} p(x_1,...,x_n) \log_2 (x_1,...,x_n)
\end{equation}
The entropy rate of the source is defined by
\begin{equation}
\label{eq:EntropyRate1}
H(\mu) = \lim_{n\rightarrow\infty} \frac{S_n(X_1,X_2,...,X_n)}{n}
\end{equation}
where $\mu$ is the measure associated with the sequence of random variables.
This limit exits when $\{X_n \}$ forms a stationary stochastic process.
Entropy rate is an important quantity in information theory as it is a measure of the average amount of 
information per symbol of a stochastic process. There is a well known formula for the entropy rate of a Markov source. 
Hidden Markov chains can be thought of as a noisy observation
Markov source emitting a sequence of symbols. Hidden Markov models have been extensively studied and the
statistical methods based on hidden Markov models have been successfully applied in diverse fields such as speech recognition,
image analysis and restoration, DNA sequencing, communication and information theory.
Even though extensive research has been carried out on hidden Markov models \cite{LR1989, EM2002}, 
the problem of deriving explicit expressions for the entropy rate of the in terms of the parameters
of the process is still an open issue. 

The entropy rate of a hidden Markov process was first studied by Blackwell in 1957 $\cite{DB1957}$ who showed that the entropy rate 
is given by the integral
\begin{equation}
\label{eq:EntropyBlackwell}
H(\mu) = \int_{\mathcal{W}} S(w) \phi(dw)  
\end{equation}
where $S$ is the Shannon entropy and $w$ belongs to the simplex $\mathcal{W}= \{w=(w_1,w_2,...,w_q):w_i \ge 0,\sum w_i=1\}$ is a $\phi(dw)$ 
is a particular measure called the Blackwell measure on the simplex $\mathcal{W}$. 
Recently there has been a renewed interest in problem of calculating the entropy rate of a hidden Markov chain.
The papers \cite{JSZ2004,HGG2006} showed a connection between the entropy rate and top Lyapunov exponent
of a product of random matrices. The study of entropy rate in the context of filtering and denoising
was done in \cite{NOW2005, OW2006}. The study of the asymptotic behavior, the smoothness and analytic 
properties and obtaining new bound and improved bounds of the entropy rate in terms of the process parameters
is carried out in \cite{OW2004,ZDKA2006,AS2009,HM2010}. Calculation of entropy rate based on ideas from statistical
mechanics is done in \cite{ZKD2005}. In this paper we follow the approach of \cite{FNS1992}, wherein a formula
similar to (\ref{eq:EntropyBlackwell}) was derived. Moreover, in \cite{FNS1992} the
support of the Blackwell measure was explicitly characterized. We see that in the case of the particular
noise model that we study in this paper the support of the measure $\phi(dw)$ simplifies and leads to an analytic
solution of the entropy rate.

In section \ref{sec:Background} we review some key results on algebraic measures
from \cite{FNS1992} that we will use in this paper. The description of the noise model
and the support of the measure $\phi(dw)$ is computed in in section \ref{sec:Noise model and supp}. The main theorem about the formula
for the entropy rate of the hidden Markov process is proved in \ref{sec:Entropy rate formulas}. Lastly in section 
\ref{sec:Entropy rate approximations} we show numerical computations of the entropy rate using approximations to the formulas obtained in section
\ref{sec:Entropy rate formulas}.

\section{Setup}
\subsection{Manifestly positive algebraic measures}
\label{sec:Background}
Let $q\in \mathbb{N}_0$ and $K=\{0,1,...,q-1\}$ and consider the chain $\Omega = K^{\mathbb{Z}}$
consisting of configurations $\omega = (...,\omega_i,\omega_{i+1},...)$ with each $w_i \in K$. Let
$\mathcal{F}_K$ be the sigma algebra generated by all the cylinder sets. Let $T:K^{\mathbb{Z}}\rightarrow K^{\mathbb{Z}}$
be the shift transformation given by
\beann
T\omega = \delta \quad \text{where} \,\, \delta_n=\omega_{n+1}
\eeann
A measure $\mu$ on $\Omega$ is called translation invariant if
\beann
\mu(E) = \mu(T^{-1}E) \quad \forall E \in \mathcal{F}_K
\eeann
In \cite{FNS1992} a special class of translation-invariant measures called algebraic measures was
constructed on the chain $\Omega$ in terms of a triplet $(U,\rho,(E_a)_{ a\in K})$ where
\begin{itemize}
\item
$U$ is a real ordered algebra (the order determined by a convex cone $U^{+}$).
\item
$\rho$ is a positive linear functional on $U$.
\item
$E_a \in U^{+}$ satisfying $\rho(AE)=\rho(EA)=\rho(A) \forall ,\, A \in U$, $\rho(E)=1$ where $E=\sum_{a\in K} E_a$. 
\end{itemize}
We state the following proposition from $\cite{FNS1992}$
\begin{proposition}
Given the triplet $(U,\rho,(E_a)_{a\in K})$ there exists a unique translation invariant probability measure $\mu$ on $K^{\mathbb{Z}}$ such that
$$ \mu(\omega_{[m,n]}) = \rho(E_{\omega_m},E_{\omega_{m+1}},...,E_{\omega_n})$$ where $\omega_{[m,n]}=(\omega_m,\omega_{m+1},...,\omega_n)$.
\end{proposition}
The measure $\mu$ was called an algebraic measure and it was  shown that if $U$ is finite dimensional the triplet $(U,\rho,(E_a)_{ a\in K})$ has a unique representation in a finite dimensional vector space. Let $(\mathbb{R}^d)^+$ denote the positive cone in $\mathbb{R}^d$ of vectors with all its components non-negative and let $(M_d)^+$  be the matrices that preserve this positive cone i.e. the all the elements of are non-negative.
An algebraic measure $\mu$ is called manifestly positive if there exists a $d \in \mathbb{N}$ positive $\tau$, $\sigma \in {(\mathbb{R}^d)}^+$ and
for each $a\in K$, a positive $E_a \in M_d^+$ such that
\beann
\mu((\omega_m,\omega_{m+1},...,\omega_n))= \ip{\tau}{E_{\omega_m} E_{\omega_{m+1}}...E_{\omega_n}\sigma }
\eeann 
Two important examples of manifestly positive algebraic measures are Markov chains and hidden Markov models.
\subsubsection{Markov chains}
Let $\{X_i\}_{i\in \mathbb{N}}$ be a Markov chain taking values in $K = \{0,...,q-1\}$ 
and having stationary measure $\mu$.
Let
\begin{itemize}
\item
$\sigma \in (\mathbb{R}^q)^+$ is the vector with all components equal to $1$.
\item 
$\tau \in (\mathbb{R}^q)^+$ such that $\tau_a = \mu((a))$ i.e. the $a^{th}$ component of the stationary distribution $\tau$.
\item
$E_a \in M_q^+ : (E_a)_{b,c} = \delta_{a,b} \frac{\mu((b,c))}{\mu((c))}$ for all $a,b,c \in K$ is the matrix with the only
non-vanishing row to be the $a^{th}$ row of the transition matrix $E = \sum_{a\in K} E_a$.
\end{itemize} Then it is easy to see that
\beann
\mu((w_m,...,w_n)) &=&\ip{\tau}{E_{w_m}...E_{w_n} 1}\\
E^*\tau            &=& \tau \\
E \sigma           &=& \sigma
\eeann
\subsubsection{Hidden Markov models}
In $\cite{FNS1992}$ a one-one correspondence was shown between functions of Markov processes (hidden Markov models)
and the class of manifestly positive algebraic measures. Let $X= \{X_i\}_{i\in \mathbb{N}}$ be a Markov chain that takes values on a finite alphabet $L$
with transition matrix $E$ stationary measure $\nu$. Let $F_a$ be the matrix with the only non-vanishing row
equal to the the $a^{th}$ of $E$. Thus
\bea
\label{eq:HMP 1}
\sum_{a \in L} F_a = E
\eea
We can represent the measure $\nu$ as in example 1  by the triplet $(\tau,1,(F_a)_{a\in L})$.
Let $Y= \{Y_i\}_{i\in \mathbb{N}}$ be a noisy observation of the Markov chain with values in $K=\{0,1,...,q-1\}$. Define the matrix $R = [r_{ab}]$ with $r_{ab}= Pr[Y_i=a|X_i=b]$ and let
\bea
\label{eq:HMP 2}
E_a = \sum_{b\in L}r_{ab} F_b
\eea
It is easy to verify that the manifestly positive representation of the stationary measure $\mu$ associated with $Y$ is $(\tau,\sigma,(E_a)_{a\in K})$.
So that
\bea
\label{eq:measure formula}
\mu((w_m,...,w_n))=\ip{\tau}{E_{w_m}...E_{w_n} \sigma}
\eea
where $\tau_a = \mu(a)$ and $\sigma$ is a vector with all components equal to 1.

\subsection{Entropy rate of manifestly positive algebraic measures}
It was shown  that under certain irreducibility conditions (see Condition 1 given below) the mean entropy rate of a manifestly positive algebraic measure can be computed as an integral with respect to a measure on the simplex $\mathcal{W}= \{w=(w_1,w_2,...,w_q):w_i \ge 0,\sum w_i=1\}$ similar to the Blackwell measure.\\
\textit{Condition 1:}\\
\begin{enumerate}[i.]
\item
There exists a $c > 0$ such that for all $a,b \in K$ $E_a E_b \geq c E_a$.
\item
There exists an $a_0 \in K$ such that the invariant subspace corresponding to the largest eigenvalue of $E_{a_0}$
is one dimensional and all other eigenvalues of $E_{a_0}$ have strictly smaller modulus.
\item
$E$ is irreducible i.e. the invariant subspace corresponding to the largest eigenvalue 1 of $E$ is one dimensional .
\end{enumerate}

We have the following theorem for the entropy rate of a function of Markov process.
\begin{theorem}[\cite{FNS1992}]
The mean entropy rate $H(\mu)$ of a manifestly positive algebraic measure $\mu$ (satisfying Condition 1) is given by
\bea
\label{eq:Entropy rate integral formula}
H(\mu)&=& \sum_{a\in K} \int_W h_a(w)\phi(dw)
\eea
where $h_a(w) = -\ip{w}{E_a\sigma}\log \ip{w}{E_a\sigma}$ and $\phi(dw)$ is a probability measure on the simplex $\mathcal{W}$.
\end{theorem}
In $\cite{FNS1992}$ an equation for $\phi(dw)$ is derived in terms of a  Markov operator $T_{\mu}$ on $\mathcal{C(W)}$, the
space of continuous functions on the simplex $\mathcal{W}$. In addition the support of the measure is 
also characterized. For functions of Markov processes Blackwell \cite{DB1957} obtained a formula similar to equation (\ref{eq:Entropy rate integral formula})
however there was no clear connection of the measure with the Markov operator and the support of the measure was also not explicitly characterized.
The Markov operator $T_{\mu}$ can be described in terms of the collection $(E_a |a\in K)$ as follows.
We first define $\Gamma_a:\mathcal{W}_0\rightarrow \mathcal{W}_0$ with $\mathcal{W}_0 = \mathcal{W} \cup \{0\}$ by  
\bea
\label{eq:Gamma defn}
\Gamma_a(\nu) =
\begin{cases}
 \frac{E_a^*\nu}{\ip{\nu E_a}{\sigma}} & \text{if} \ip{\nu E_a}{\sigma} \ne 0 \\
 0 & \text{otherwise}
\end{cases}
\eea
Let $\mathcal{C}_0(\mathcal{W}_0)$ be the set of continuous real valued functions that vanish on $0$, 
then $T_{\mu}: \mathcal{C}_0(\mathcal{W}_0) \rightarrow \mathcal{C}_0(\mathcal{W}_0)$ is defined by
\bea
\label{eq:Markov operator defn}
(T_{\mu} f)(w) &=& \sum_{a \in K} \ip{w}{E_a \sigma} f(\Gamma_a w)
\eea
By the Riesz-Markov theorem (Theorem IV. 14 \cite{RS1980}) there is a one to one correspondence between the space
of measures and normalized positive linear functionals on $\mathcal{C(W)}$. Therefore
\bea
\phi(f) = \int_{\mathcal{W} } f(w) \phi(dw) \quad f \in \mathcal{C(W)}
\eea
In \cite{FNS1992} the measure $\phi(dw)$ was characterized as the unique measure on $\mathcal{C(W)}$ that is invariant under $T_{\mu}$ i.e.
\bea
\label{eq:Invariant Measure}
\phi(f)&=&\phi(T_{\mu} f) \, f \in \mathcal{C(W)}
\eea
The support of $\phi(dw)$ is given by
\begin{equation}
\label{eq:support}
\text{supp}(\phi) = \Delta \equiv \overline{\{\Gamma_{(\omega)}\bar{\tau} | \omega \in K^n, n\in N\}}
\end{equation}
where $\Gamma_{(\omega)} = \Gamma_{\omega_{n-1}}\cdots \Gamma_{\omega_0}$ and $\bar{\tau}$ is the only non-trivial fixed point of $\Gamma_{a_0}$.

\section{Noise model and support of the measure $\phi$}
\label{sec:Noise model and supp}
\begin{figure}[ht]
\begin{center}
\epsfig{file=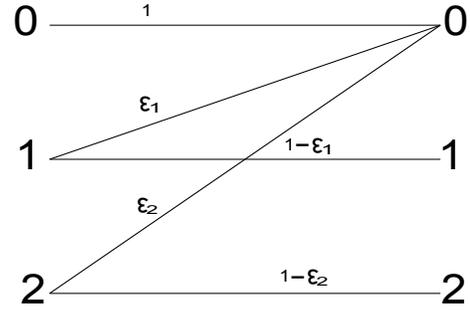,height=4cm,width=6cm}
\caption{The noise model for $q=3$. If 0 is transmitted 0 is received with probability 1. If 1 of 2 are transmitted
there is an error probability of $\epsilon_1$ and $\epsilon_2$ respectively of the received symbol being 0.}
\label{fig:Model}
\end{center}
\end{figure}

The entropy rate formulas derived in this paper are for a general $q$ state hidden Markov process described by a particular noise model.
In this section we describe the noise model and the support of the measure $\phi(dw)$ given by equation (\ref{eq:support}) for this noise model.
The noise model that we work assumes that noise does not affect exactly one of the input symbols, say the symbol $0$. If the symbol $0$ is
transmitted then it is always unambiguously received at the other end. On the other hand if any of the other symbol
is transmitted then it is either received without any error or received as the symbol $0$ with a small error probability.
That is $P(Y_i=0|X_i=0) =1$, $P(Y_i=0|X_i=a) = \epsilon_a$ and $P(Y_i=a|X_i=a) = 1- \epsilon_a$ for $a=1,...,q-1$ and $P(Y_i=b|X_i=a)=0$ when $0\neq b \neq a$. See figure \ref{fig:Model} for a description of the model in the special case of $q=3$. In this paper without loss of generality we shall assume that the unambiguous symbol is $0$.
Let the matrices $\{F_a\}$ be the matrices that describe the uncorrupted Markov source as in equation (\ref{eq:HMP 1}).
For this noise model we write the matrices $\{E_a\}$ given by equation (\ref{eq:HMP 2}) as
\bea
\label{eq:Matrix E_a defn}
E_0 &=& F_0 + \sum_{a=1}^{q-1} \epsilon_a F_a\\ \nonumber
E_a &=& (1-\epsilon_a) F_a \quad \text{for} \,a= 1,...,q-1\\ \nonumber
\sum_{a\in K} E_a &=& E
\eea
Let $e_i$ denote the transpose of the $(i-1)^{st}$ row of $E=[e_{ij}]$. Let $p = \min_{ij} e_{ij}$ and $P=\max_{ij} e_{ij}$.
Our only assumption will be\\

\textit{Assumption} 1:\\
$0<p \le P < 1$, $\epsilon_0 = 1$ and $\epsilon_a > 0 \quad \forall a \in \{1,...,q-1\}$.\\

From equation (\ref{eq:Gamma defn}) one can see that for each $\nu \in \mathcal{W}$
\bea
\label{eq:Gamma_0}
\Gamma_0 \nu &=& \frac{\sum_{a=0}^{q-1}\epsilon_a\nu_a \mathbf{e_a} }{\sum_{a=0}^{q-1}\epsilon_a\nu_a} = \sum_{a=0}^{q-1}\alpha_a \mathbf{e_a}
\eea
$\sum_{a=0}^{q-1} \alpha_a = 1$ with $\alpha_a = \frac{\epsilon_a \nu_a}{\sum_{a=0}^{q-1}\epsilon_a \nu_a}$.\\
Because of \textit{Assumption} 1 we get
\bea
\label{eq:Gamma_0 bound}
p  \le (\Gamma_0 \nu)_a \le P \quad  \forall a \in K.
\eea
Also one gets from equation (\ref{eq:Gamma defn}) for all $a \in \{1,...,q-1\}.$
\bea
\label{eq:Gamma_a}
\Gamma_a \nu  = 
\begin{cases}
e_a  & \text{if} \, \nu_a \ne 0\\
0 & \text{otherwise}
\end{cases}
\eea
A key factor that simplifies our analysis of the entropy rate is that the support of $\phi(dw)$ given by equation (\ref{eq:support}) simplifies significantly with this noise model. We have the following proposition about the
support of the measure $\phi$. 
\begin{proposition}
\label{prop:support of phi}
The support of the measure $\phi$ is given by
\bea
\Delta = \overline{\{\Gamma_0^m e_j | j \in \{1,..,q-1\} ; m \in \mathbb{N}_0\}}
\eea
\end{proposition}
\IEEEproof
We know from equation (\ref{eq:support}) that the support of the measure $\phi$ is given by
\beann
\Delta =  \overline{\{\Gamma_{(\omega)}\bar{\tau} | \omega \in K^n, n\in N\}}
\eeann
where $\bar{\tau}$ is the non-trivial fixed point of $\Gamma_0$. 
Since by equation (\ref{eq:Gamma_0 bound}) $\bar{\tau}  \in (\mathbb{R}^q)^+$, it follows from equation (\ref{eq:Gamma_a}) that $\Gamma_a \bar{\tau} = e_a$ and
$\Gamma_a e_a = e_a$ for $a \ne 0$. Therefore 
\beann
\Delta =  \overline{\{\bar{\tau}\} \cup \{\Gamma_0^m e_j | j \in \{1,..,q-1\}; m\in \mathbb{N}_0\}}
\eeann
One can show (\cite{FNS1992} equation (28)) that for any $\nu \in \mathcal{W}^+$ there exists constants $C_1$ and $\rho <1$ such that
\bea
\label{eq:Norm estimate of tau}
\|\Gamma_0^{k+1}\nu - \bar{\tau}\|_1 \leq C_1 \rho^k
\eea
This implies that all subsequences in $\Delta$ converge to $\bar{\tau}$ and 
hence $\bar{\tau}$ is the only accumulation point in $\Delta$. Therefore
\beann
\Delta &=& \overline{\{\Gamma_0^m e_j | j \in \{1,..,q-1\} ; m \in \mathbb{N}_0\}}\\
\eeann
Moreover for any $j\in \{1,...,q-1\}$
\bea
\label{eq:Convergence to tau}
\lim_{m\rightarrow\infty} \Gamma_0^m e_j   &=&  \bar{\tau} 
\eea
\IEEEQED 

We note that $\Delta$ may consist finite or infinitely many distinct points in $\mathcal{W}$. For instance in the case if $e_0 = e_1=\cdots=e_{q-1}$
then it is easy to see that $\bar{\tau} = e_0 = \cdots = e_{q-1}$ and $\Delta$ consists of a single element i.e. $\Delta = \{e_0\}$. It is clear from (\ref{eq:Norm estimate of tau}) that if $\Delta$ is a finite set then $\bar{\tau} = \Gamma_0^n e_j$ for some $n \in N$ and some $j\in \{1,...,q-1\}$. On the other hand $\Delta$ can be countably infinite as can be seen from the following lemma.

\begin{lemma}
\label{lem:Gamma_0 one-one}
If $E_0^*$ is one to one then $\Delta$ is countably infinite set of distinct elements.
\end{lemma}
\IEEEproof
We will first show that $E_0^*$ is one-one $\implies$ $\Gamma_0:\mathcal{W}\rightarrow \mathcal{W}$ is one-one.
\beann
\text{If} \quad \Gamma_0 \mathbf{\nu} &=& \Gamma_0 \eta \quad \text{for} \, \mathbf{\nu},\mathbf{\eta} \in \mathcal{W}\\
\text{then} \quad \frac{E_0^*\mathbf{\nu}}{\ip{\nu}{E_0 \sigma}} &=& \frac{E_0^*\mathbf{\eta}}{\ip{\eta}{E_0 \sigma}}\\
\text{Therefore} \quad E_0^*(\mathbf{\nu} - C \mathbf{\eta}) &=& 0
\eeann
where $C = \frac{\ip{\nu}{E_0 \sigma}}{\ip{\eta}{E_0 \sigma}}$.
But $E_0^*$ is one-one implies
\beann
\mathbf{\nu} = C\mathbf{\eta}
\eeann
and since $\mathbf{\nu}, \mathbf{\eta} \in \mathcal{W}$ therefore $C =1$ so $\mathbf{\nu} = \mathbf{\eta}$.
Since $E_0^*$ is one-one the vectors $e_a \in \mathcal{W}, \, a\in K$ form a linearly independent
set. We will show that all the elements of $\Delta$ are distinct i.e. for any $m,n \geq 0$
\bea
\label{eq:Gamma_0 one-one}
\Gamma_0^m e_i  &\neq&  \Gamma_0^n e_j\\
\Gamma_0^m e_i  &\neq& \bar{\tau} \nonumber
\eea

Assume $\Gamma_0^m e_i = \Gamma_0^n e_j$ for some $m,n \geq 0$ and for $i,j \in \{1,...,q-1\}$,
If $m$ and $n$ are both zero then $e_i = e_j$ but this contradicts the fact
that $e_j$'s form a linearly independent set. Assume wlog that $m > n$. Since
$\Gamma_0$ is one-one we arrive at $\Gamma_0^k e_i = e_j$ for $k = m-n$.
Again using equation (\ref{eq:Gamma_0}) we arrive at a contradiction that $e_j$'s form a linearly independent set.

From above,  $\Gamma_0 e_a \neq e_a$ for any $a \in K$ so $e_a \neq \bar{\tau}$.
If $\Gamma_0 ^m e_a = \bar{\tau}$ for some $m \geq 1$ then 
$\Gamma_0 (\Gamma_0^{m-1} e_a) = \bar{\tau}$. Since $\bar{\tau}$ is a fixed point of $\Gamma_0$
and $\Gamma_0$ is one to one we conclude that $(\Gamma_0^{m-1}) e_a = \bar{\tau}$.
Repeating the argument we get $e_a = \bar{\tau}$ which again is a contradiction.
\IEEEQED 

\begin{lemma}
If $\Delta$ is countably infinite then $\phi(\bar{\tau}) = 0$
\end{lemma}
\IEEEproof
Since $\Delta$ is countably infinite there exists a $j \in \{1,...,q-1\}$ such that $\lim_{m\rightarrow\infty} \Gamma_0^m e_j = \bar{\tau}$.
Consider the set
\beann
\Delta_{j,m} = \overline{\{\Gamma_0^k e_j | k \geq m\}}
\eeann
for $m\in \{1,2,...\}$. This is a decreasing sequence of sets with
\beann
\bigcap_{m \in \mathbb{N}}\Delta_{j,m} &=& \{\bar{\tau}\}\\
\text{Therefore} \quad \phi(\bar{\tau}) &=& \lim_{m\rightarrow\infty} \phi(\Delta_{j,m})
\eeann
Let $f_m \in \mathcal{C}({\mathcal{W}})$ be defined so that 
\beann
f_m(\nu) =
\begin{cases}
1 & \nu \in \Delta_{j,m}\\
0 & \text{otherwise}
\end{cases}
\eeann
So,
\beann
\phi(\Delta_{j,m}) &=& \int_{\Delta} f_m(\nu) d\phi
\eeann
By (\ref{eq:Markov operator defn}) and (\ref{eq:Invariant Measure}) we get
\beann
\phi(\Delta_{j,m}) &=& \int_{\Delta} \sum_{a\in K}\ip{\nu}{E_a \sigma} f_m(\Gamma_a\nu) d\phi
\eeann
Therefore we get
\beann
\phi(\Delta_{j,m}) &=& \int_{\Delta_{m-1}}\ip{\nu}{E_0 \sigma} d\phi
\eeann
where we used $(\ref{eq:Gamma_a})$ and the fact that $\Gamma_0\nu \in \Delta_m$ only if $\nu \in \Delta_{m-1}$.
We have
\beann
\phi(\Delta_{j,m}) &=& \int_{\Delta_{m-1}}\ip{\nu}{E_0 \sigma} d\phi\\
                   &=& \int_{\Delta_{m-1}} \sum_{a=0}^{q-1} \epsilon_a \nu_a d\phi
\eeann
Since $\sum_a \nu_a = 1$, $\epsilon_0 =1$  and $\epsilon_a < 1$ for $a \neq 0$
\beann
r = \sum_{a=0}^{q-1} \epsilon_a \nu_a < 1.
\eeann
So,
\beann
\phi(\Delta_{j,m}) &\leq& r \phi(\Delta_{j,m-1}) \implies \\
\phi(\Delta_{j,m}) &\leq& r^{m-1} \phi(\Delta_1)
\eeann
Therefore,
\beann
\phi(\bar{\tau}) &=& \lim_{m\rightarrow\infty}\phi(\Delta_{j,m}) \leq \lim_{m\rightarrow\infty} r^{m-1} \phi(\Delta_1) = 0.
\eeann
\IEEEQED

The next lemma shows that our \textit{Assumption} 1 on the Markov transition matrix and the noise parameters is enough  to satisfy \text{Condition} 1.
\begin{lemma}
\label{lem:Condition 1}
If $0<p \leq P < 1$ and $e_a > 0$ for all $a\in K$ then the matrices $(E_a)_{a\in K}$ satisfy \textit{Condition 1}.
\end{lemma}
\IEEEproof
Condition 1 i. can be verified by a simple computation by choosing $c= \epsilon p^{q-1}$ where $\epsilon = \min_i \epsilon_i$. 
ii. and iii. follow from the Perron-Frobenius theorem and the fact
that $E$ is a Markov transition matrix.
\IEEEQED 

\section{Entropy rate formulas for a hidden Markov process}
\label{sec:Entropy rate formulas}
In this section we apply the results on manifestly positive algebraic measures in \ref{sec:Background}
to the case of the hidden Markov model described in section \ref{sec:Noise model and supp}. 
We divide our support set $\Delta$ into disjoint sets in the following way
\beann
\Delta_1 &=& \{\Gamma^m_0 e_1: m\in \mathbb{N}_0\}\\
\Delta_2 &=& \{\Gamma^m_0 e_2: m\in \mathbb{N}_0\} \backslash \{\Delta_1\}\\
\Delta_3 &=& \{\Gamma^m_0 e_3: m\in \mathbb{N}_0\} \backslash \{\Delta_1 \cup \Delta_2\} \\
\hdots   &\hdots & \quad \hdots \quad \hdots \quad \hdots \quad \hdots \quad \hdots \\
\Delta_{q-1} &=& \{\Gamma_0^m e_{q-1}: m\in \mathbb{N}_0\} \backslash \{\bigcup_{i=1}^{q-2}\Delta_i\}
\eeann  
Define
\bea
\label{eq:c_jm}
c_{j,m} = \prod_{i=1}^m \ip{\Gamma_0^{m-i} e_j}{E_0 \sigma}
\eea
Let $A$ be the $q\times q-1$ matrix defined by entries.\\
If $i\ne q$ then
\bea
\label{eq:A matrix 1}
A_{ij} =
\begin{cases}
-\delta_{ij} + \sum_{m=0}^{|\Delta_j|}\ip{\Gamma_0^m e_j}{E_i \sigma}c_{j,m}& q\neq 2 \\
0 & \text{if} \, q=2 
\end{cases}
\eea
\bea
\label{eq:A matrix 2}
A_{qj} &=& \sum_{m=0}^{|\Delta_j|} c_{j,m}
\eea
\bea
\label{eq:Phi and b}
\Phi &=& [\phi(e_1),...,\phi(e_{q-1})]^{'} \in \mathbb{R}^{q-1}\\
 b   &=& [0,0,...,1]^{'} \in \mathbb{R}^q \nonumber
\eea
\begin{theorem}
Under \textit{Assumption} 1 the entropy rate of the measure $\mu$ associated with the hidden Markov process with the noise model described in section \ref{sec:Noise model and supp} is given by
\bea
\label{eq:Entropy rate formula}
H(\mu) = \sum_{j=1}^{q-1}\sum_{m=0}^{|\Delta_j|}\sum_{a=0}^{q-1} h_a(\Gamma_0^m e_j) c_{j,m} \Phi_j 
\eea
where $\Phi_j$ is the $j^{th}$ coordinate of the solution to the matrix equation $A\Phi = b.$
\end{theorem}
\IEEEproof
Let 
\beann
f_{j,m}(\nu)=
\begin{cases}
1 & \text{if} \, \nu = \Gamma_0^m e_j\\
0 & \text{otherwise}
\end{cases}
\eeann 
for $m\in \mathbb{N}$ and $j=1,...,q-1$. We have
\beann
\phi(\Gamma_0^m e_j) &=& \int_{\mathcal{W}}f_{j,m} d\phi
\eeann
By (\ref{eq:Markov operator defn}) and (\ref{eq:Invariant Measure})  we get
\beann
\phi(\Gamma_0^m e_j)  &=& \int_{\mathcal{W}} \sum_{a\in K} \ip{\nu}{E_a \sigma}f_{j,m} (\Gamma_a \nu) d\phi \\
\eeann
But $\Gamma_a(\nu) = e_a$  and $\Gamma_0 \nu = \Gamma_0^m e_j$  only if $\nu = \Gamma_0^{m-1} e_j$ therefore
\beann
\phi(\Gamma_0^m e_j) &=& \ip{\Gamma_0^{m-1}e_j}{E_0 \sigma} \phi(\Gamma_0^{m-1} e_j)
\eeann
and iterating we get
\bea
\label{eq:Gamma_0 iteration}
\phi(\Gamma_0^m e_j) &=& c_{j,m} \phi(e_j)\\
\text{where} \quad c_{j,m} &=& \prod_{i=1}^m \ip{\Gamma_0^{m-i}e_j}{E_0 \sigma} \nonumber
\eea
Similarly we use the functions $f_{j,0}$ to solve for $\phi(e_1),...\phi(e_{q-1})$.
\beann
\phi(e_i) &=& \int_{\mathcal{W}} \sum_{a\in K}\ip{\nu}{E_a\sigma} f_{i,0}(\Gamma-a\nu) d\phi\\
&=& \sum_{\nu\in \Delta} \ip{\nu}{E_i\sigma} \phi(\nu)\\
&=& \sum_{j=1}^{q-1} \sum_{m=0}^{|\Delta_j|} \ip{\Gamma_0^m e_j}{E_i \sigma}\phi(\Gamma_0^m e_j)
\eeann
Using $(\ref{eq:Gamma_0 iteration})$ we get
\bea
\label{eq:Linear system 1}
\phi(e_i) &=& \sum_{j=1}^{q-1}\phi(e_j) \sum_{m=0}^{\infty} \ip{\Gamma_0^m e_j}{E_i \sigma} c_{j,m}
\eea
Now the measure of the whole set $\phi(\Delta) = 1$ so
\bea
\label{eq:Linear system 2}
\sum_{j=1}^{q-1} \sum_{m=0}^{|\Delta_j|} \phi(\Gamma^m e_j) &=& \sum_{j=1}^{q-1}\phi(e_j)\sum_{m=0}^{\infty}c_{j,m} =1
\eea
Equations (\ref{eq:Linear system 1}) and (\ref{eq:Linear system 2}) form a system of $q$ linear equations
$A\Phi = b$ with $A \in M_{q,q-1}$, $\Phi \in \mathbb{R}^{q-1}$ and $b\in \mathbb{R}^q$ as given by 
(\ref{eq:A matrix 1}),(\ref{eq:A matrix 2}),(\ref{eq:Phi and b}). By lemma (\ref{lemma:c_jm bound 1})
$\sum_{m=0}^{|\Delta_j|} c_{j,m} < \infty$ which ensures that each $A_{ij}$ is bounded.
Now from the integral formula for the entropy rate given by (\ref{eq:Entropy rate integral formula}) and the support of the 
measure given by proposition (\ref{prop:support of phi}) and we get that
\beann
H(\mu) &=& \sum_{j=1}^{q-1} \sum_{m=0}^{|\Delta_j|} \sum_{a=0}^{q-1} h_a(\Gamma_0^m e_j)\phi (\Gamma_0^m e_j)
\eeann
From (\ref{eq:Gamma_0 iteration}) we get	
\beann
H(\mu) &=& \sum_{j=1}^{q-1} \sum_{m=0}^{|\Delta_j|} \sum_{a=0}^{q-1} h_a(\Gamma_0^m e_j)c_{j,m} \phi(e_j)\\
H(\mu) &=& \sum_{j=1}^{q-1} \sum_{m=0}^{|\Delta_j|} \sum_{a=0}^{q-1} h_a(\Gamma_0^m e_j)c_{j,m} \Phi_j
\eeann
\IEEEQED

For the case when $E_0^*$ is one-one from lemma \ref{lem:Gamma_0 one-one} and equation (\ref{eq:Gamma_0 one-one}) it follows that
\beann
H(\mu) &=& \sum_{j=1}^{q-1} \sum_{m=0}^{\infty} \sum_{a=0}^{q-1} h_a(\Gamma_0^m e_j)c_{j,m} \Phi_j
\eeann

\begin{lemma}
\label{lemma:c_jm bound 1}
For all $j=1,...,q-1$ and $m\in \mathbb{N}_0$  
\bea
\label{eq:gamma defn}
c_{j,m} &<& \gamma^m \quad \text{where} \nonumber\\
\gamma   &=& \max_{j} \sup_{k} \sum_{a=0}^{q-1} \epsilon_a [\Gamma_0^k e_j]_a < 1
\eea

\end{lemma}
\IEEEproof
\bea
\label{eq:c_jm equation}
c_{j,m} &=& \prod_{i=1}^m \ip{\Gamma_0^{m-i} e_j}{E_0 \sigma}\\ \nonumber
        &=& \prod_{i=1}^m \sum_{a=0}^{q-1}\epsilon_a[\Gamma_0^{m-i} e_j]_a
\eea
We know by (\ref{eq:Gamma_0 bound}) $p \leq [\Gamma_0^k e_j]_a \leq P$ for each $a\in K$.
Since $\Gamma^k e_j \in \mathcal{W}$, $\sum_{a\in K}  [\Gamma^k e_j]_a =1$.
Now, by \textit{Assumption} 1. 
$\epsilon_a < 1$ if $a\in \{1,..,q-1\}$ and  $\epsilon_0 = 1.$ 
\beann
\text{So} \quad \sum_{a=0}^{q-1} \epsilon_a [\Gamma^k e_j]_a < 1 \quad \forall k \in \mathbb{N}_0
\eeann
Moreover by (\ref{eq:Convergence to tau}) 
\beann
\lim_{k\rightarrow\infty} \Gamma_0^k e_j = \bar{\tau} \in \mathcal{W}^+
\eeann
Therefore,
\beann
\sup_{k} \sum_{a=0}^{q-1} \epsilon_a [\Gamma_0^k e_j]_a < 1
\eeann
and
\beann
\gamma = \max_{j} \sup_{k} \sum_{a=0}^{q-1} \epsilon_a [\Gamma_0^k e_j]_a < 1
\eeann
\IEEEQED

We also prove the following bound for $c_{jm}$ which will be useful in the numerical estimates in section \ref{sec:Entropy rate approximations}.
\begin{lemma}
\label{lemma:c_jm bound 2}
Let $\epsilon = \max_{a \in \{1,..,q-1\}} \epsilon_a$. If $\epsilon < p$ then
\bea
\label{eq:r defn}
c_{j,m}  &<& r^m  \quad \text{where}\nonumber\\
r        &=& 1 - (q-1)(p-\epsilon P) < 1
\eea
\end{lemma}
\IEEEproof
$\epsilon < p \implies r=(1-(q-1)p +(q-1)\epsilon P) < 1.$
\beann
c_{j,m} &=&     \prod_{i=1}^m \ip{\Gamma_0^{m-i} e_j}{E_0 \sigma}\\
        &=&     \prod_{i=1}^m \big[\Gamma_0^{m-i} e_j \big ]_0 + \sum_{a=1}^{q-1} \epsilon_a\big [\Gamma_0^{m-i} e_j \big]_a\\
        &\leq & \prod_{i=1}^m  (1-(q-1)p +(q-1)\epsilon P)\\
        &=& r^m
\eeann
\IEEEQED

\section{Approximating the entropy rate}
\label{sec:Entropy rate approximations}
In this section we present some numerics approximations to the entropy rate formulas
that were derived in section \ref{sec:Entropy rate formulas}. We show that if we
take only the first $N$ terms of the matrix $A$ given by (\ref{eq:A matrix 1}) and (\ref{eq:A matrix 2})
for the entropy rate calculations  then this gives an approximation of order $O(\gamma^{N+1})$ where is given by (\ref{eq:gamma defn}).

Let
\bea 
\label{eq:A matrix approximation}
A=\hat{A} + R
\eea
where the entries of $R$ are the $N^{th}$ tails of the entries of $A$. 
Let $\hat{\Phi}$ be the least square solution to 
\bea
\label{eq:Linear system}
\hat{A} \Phi = b
\eea
Therefore
\bea
\label{eq:Least squares solution}
\hat{\Phi} = \hat{A}^{\dagger} b
\eea
where $\hat{A}^{\dagger}$ is the pseudo-inverse of $A$.
We first prove the following lemma
\begin{lemma}
\label{lem:Phi 1-Norm}
\beann
\|\Phi - \hat{\Phi}\|_1 \leq \frac{q\|\hat{A}^{\dagger}\|_1}{1-\gamma}
\eeann
\end{lemma}
\IEEEproof
From (\ref{eq:Linear system}) and (\ref{eq:A matrix approximation})
\beann
A\Phi &=& \hat{A}\Phi + R \Phi = b\\
\eeann
Substituting  (\ref{eq:Least squares solution})
\beann
\hat{A}\Phi -\hat{A}\hat{\Phi} &=& b -R\Phi - \hat{A}\hat{\Phi}\\
\hat{A}(\Phi -\hat{\Phi} )&=& b-R\Phi -\hat{A}\hat{\Phi} \implies 
\eeann
\beann
\|\Phi-\hat{\Phi}\|_1 &=& \|\hat{A}^{\dagger}b-\hat{A}^{\dagger}R\Phi -\hat{\Phi}\|\\
\|\Phi-\hat{\Phi}\|_1 &\leq& \|\hat{A}^{\dagger}\|_1\|R\Phi\|_1
\eeann
But since $R$ consists of the tail of the entries each entry in $R$ is bounded by 
$\sum_{m=N+1}^{\infty}\gamma^m = \frac{\gamma^{N+1}}{1-\gamma}$.
\beann
\|R\Phi\| \leq \frac{\gamma^{N+1}}{1-\gamma}\left\|\begin{pmatrix} 1 &\cdots& 1 \\ &\ddots& \\ 1& \cdots& 1  \end{pmatrix}\Phi \right \|_1 \leq \frac{\gamma^{N+1}q}{1-\gamma}
\eeann
\beann
\|\Phi-\hat{\Phi}\|_1 \leq \|\hat{A}^{\dagger}\|_1\|R\Phi\|_1 \leq \frac{q\|\hat{A}^{\dagger}}{1-\gamma} \gamma^{N+1}
\eeann
\IEEEQED

\begin{theorem}
\label{thm:Entropy rate appx}
Under \textit{Assumption} 1 the entropy rate of the measure $\mu$ associated with the hidden Markov process 
described in section \ref{sec:Noise model and supp} can be approximated to $O(\gamma^{N+1})$ by
\beann
H_N(\mu)= \sum_{j=1}^{q-1} \sum_{m=0}^N\sum_{a=0}^{q-1} h_a(\Gamma_0^m e_j) c_{j,m} \hat{\Phi}_j
\eeann
\end{theorem}
\IEEEproof
\beann
\text{err}(N) &=& |H(\mu) - H_N(\mu)|
\eeann
\begin{align*}
\text{err}(N)     &= \big| \sum_{j=1}^{q-1} \sum_{m=N+1}^{\infty}\sum_{a=0}^{q-1} h_a(\Gamma_0^m e_j) c_{j,m} \hat{\Phi}_j +\\
           &\quad \sum_{j=1}^{q-1} \sum_{m=0}^{\infty}\sum_{a=0}^{q-1} h_a(\Gamma_0^m e_j) c_{j,m}(\Phi_j - \hat{\Phi}_j)\big|\\
           &\leq q\sum_{j=1}^{q-1} \sum_{m=N+1}^{\infty} \gamma^m \hat{\Phi}_j + q\sum_{j=1}^{q-1}\sum_{m=0}^{\infty}\gamma^m|\Phi_j-\hat{\Phi}_j|\\
           &= q\frac{\gamma^{N+1}}{1-\gamma} \|\hat{\Phi}\|_1 + q\frac{1}{1-\gamma}\|\Phi-\hat{\Phi}_j\|_1\\
           &\text{By lemma \ref{lem:Phi 1-Norm}}\\
           &\leq q\frac{\gamma^{N+1}}{1-\gamma} + q\frac{1}{1-\gamma}\big(\frac{q\|\hat{A}^{\dagger}\|_1}{1-\gamma}\gamma^{N+1}\big)
\end{align*}
Therefore

\bea
\label{eqn:err_N bound1}
\text{err}(N) &\leq& q\frac{\gamma^{N+1}}{1-\gamma}\big(1+\frac{q\|\hat{A}^{\dagger}\|_1}{1-\gamma}\big)\\
               &=&  O(\gamma^{N+1})\nonumber
\eea
\IEEEQED

Next, we present a numerical example for approximating the entropy rate formulas given by theorem \ref{thm:Entropy rate appx}.
We can get estimates on $\text{err}(N)$ using the bound (\ref{eqn:err_N bound1}), but $\gamma$ is difficult to compute.
However, note that lemma \ref{lemma:c_jm bound 2} implies that whenever $\epsilon < p$ the bound
\bea
\label{eqn:err_N bound2}
\text{err}(N) &\leq& q\frac{r^{N+1}}{1-r}\big(1+\frac{q\|\hat{A}^{\dagger}\|_1}{1-r}\big)
\eea
is also obtained where $r$ is given by (\ref{eq:r defn}).
In example that follows, in order to estimate $\text{err}(N)$
we work with an additional assumption that $\epsilon < p$.

\subsubsection{Example 1}
\begin{figure}[ht]
\begin{center}
\epsfig{file=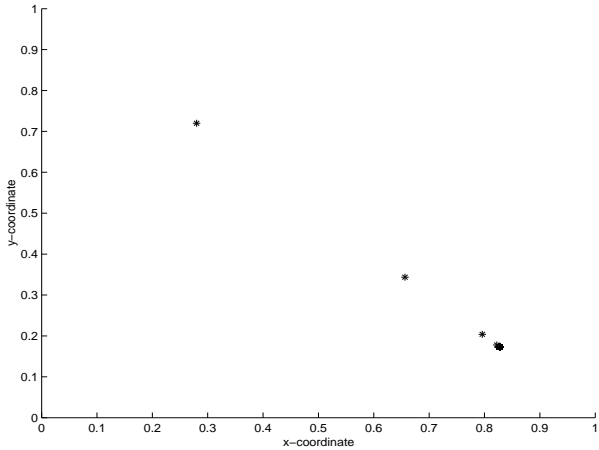,height=6cm,width=8cm}
\caption{The support of the measure $\bar{\phi}$. It is seen that the support converges to accumulation point $\bar{\tau}$.}
\label{fig:Entsupp}
\end{center}
\end{figure}

In the example we let $q=2$, $\epsilon = 0.01$. The transition matrix we use is
\beann
E &=& \begin{pmatrix} 0.85 & 0.15  \\ 0.28 & 0.72 \end{pmatrix}
\eeann
Table \ref{table:entropy rate 1} shows the estimate entropy rate $H_N(\mu)$ using
the formula given by theorem \ref{thm:Entropy rate appx}. Figure \ref{fig:Entsupp} shows
the support of the measure $\bar{\phi}$ The value of $\epsilon$ is chosen to be $0.2$ so as to make the plot clearly visible.
\begin{table}[ht]
\begin{center}
\begin{tabular}{|c|c|c|}	
\hline
N       &   $H_N(\mu)$       & err(N) bound\\
\hline
10      &   0.71399868740464 & 15.6656\\
\hline
20      &   0.70277846315804 & 3.4068 \\
\hline   
30      &   0.70083402087899 & $0.7408$\\
\hline
40      &   0.70045844593354 & $0.1611$\\
\hline
50      &   0.70038443295765 & $0.0350$\\  
\hline
60      &   0.70036979023825 & $0.0076$\\
\hline
70      &   0.70036689107994 & $0.0016$\\
\hline   
80      &   0.70036631697843 & $3.6038 \times 10^{-4}$\\
\hline
90      &   0.70036620328938 & $7.837 \times 10^{-5}$\\
\hline
100      &  0.70036618077546 & $1.7044 \times 10^{-5}$\\  
\hline
\end{tabular}
\caption{List of values of entropy rate for $q = 2$}
\label{table:entropy rate 1}
\end{center}
\end{table}

\subsubsection{Example 2}
\begin{figure}[ht]
\begin{center}
\epsfig{file=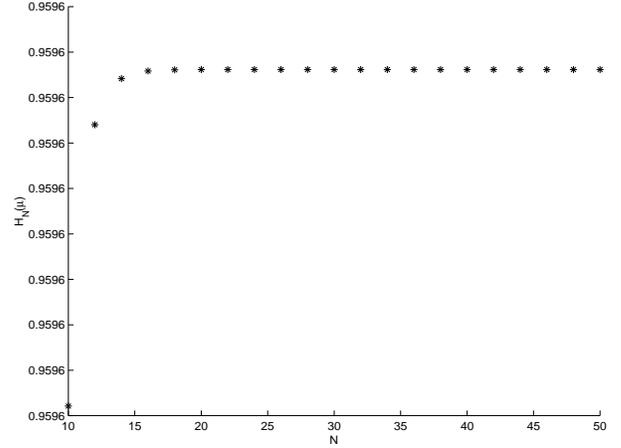,height=6cm,width=8cm}
\caption{Plot of Entropy rate $H_N(\mu)$ computed using new formulas versus N. The estimate converges very quickly with N.}
\label{fig:Ent3d}
\end{center}
\end{figure}

\begin{figure}[H]
\begin{center}
\epsfig{file=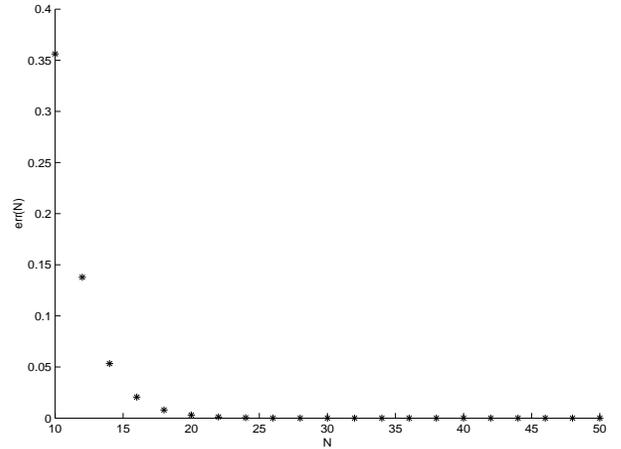,height=6cm,width=8cm}
\caption{Plot of the error $\text{err}(N)$ versus N. A reasonable bound on the error is obtained with fewer than 50 terms of the sum.}
\label{fig:Enterr3d}
\end{center}
\end{figure}

In the example we let $q=3$, $\epsilon_1 = 0.01$ and $\epsilon_2 = 0.02$. The transition matrix we use is
\beann
E &=& \begin{pmatrix} 0.4 & 0.25 & 0.35 \\ 0.25 & 0.45 & 0.3\\ 0.2 & 0.55 & 0.25 \end{pmatrix}
\eeann
Table \ref{table:entropy rate 2} shows the estimate entropy rate $H_N(\mu)$ using
the formula given by theorem \ref{thm:Entropy rate appx}. Figure \ref{fig:Ent3d} show a plot of the 
entropy rate $H_N(\mu)$ versus $N$. It is seen that the formulas for $H_N(\mu)$ converge very quickly with $N$.
Figure \ref{fig:Enterr3d} shows the plot of the bound on $\text{err}(N)$ given by equation (\ref{eqn:err_N bound2}) versus $N$.
It can be seen that we get a very good bound on $\text{err}(N)$ within a few terms. 

\begin{table}[ht]
\begin{center}
\begin{tabular}{|c|c|c|}	
\hline
N       &   $H_N(\mu)$       & err(N) bound\\
\hline
10      &   0.95961052113515 & 0.3561\\
\hline
20      &   0.95961126155225 & 0.0030\\
\hline   
30      &   0.95961126164043 & $2.6758 \times 10^{-5}$\\
\hline
40      &   0.95961126164044 & $2.3193 \times 10^{-7}$\\
\hline
50      &   0.95961126164044 & $2.0103 \times 10^{-9}$\\  
\hline
\end{tabular}
\caption{List of values of entropy rate for $q = 3$}
\label{table:entropy rate 2}
\end{center}
\end{table}

\bibliographystyle{IEEEtran}
\bibliography{ER_paper}

\end{document}